\documentclass[lettersize,journal]{IEEEtran}
\usepackage{amssymb}
\usepackage{algorithmic}
\usepackage{algorithm}
\usepackage{hyperref}
\usepackage{array}
\usepackage{textcomp}
\usepackage{stfloats}
\usepackage{url}
\usepackage{verbatim}
\usepackage{graphicx}
\usepackage{booktabs}
\usepackage{amsmath}
\usepackage{array}
\usepackage{xcolor}
\usepackage{cite}
\usepackage{multirow}
\usepackage{bm}
\usepackage{pgfplots}
\usepackage{tikz}
\usepackage{amsthm}

\usetikzlibrary{arrows.meta}

\DeclareMathOperator*{\argmax}{arg\,max}
 
\definecolor{c11}{rgb}{0.4, 1.0, 1.0}
\definecolor{c12}{rgb}{0.3, 0.75, 0.9}
\definecolor{c13}{rgb}{0.2, 0.5, 0.8}
\definecolor{c14}{rgb}{0.1, 0.25, 0.7}
\definecolor{c15}{rgb}{0.0, 0.0, 0.6}

\definecolor{c21}{rgb}{1, 0.5, 0.5}
\definecolor{c22}{rgb}{1, 0.4, 0.4}
\definecolor{c23}{rgb}{1, 0.3, 0.3}
\definecolor{c24}{rgb}{1, 0.2, 0.2}
\definecolor{c25}{rgb}{1, 0.1, 0.1}

\definecolor{c31}{rgb}{0.0, 0.6, 1.0}
\definecolor{c32}{rgb}{0.1, 0.4, 0.7}
\definecolor{c33}{rgb}{0.2, 0.2, 0.4}

\begin{document}

\title{Viterbi State Selection for Discrete Pinching Antenna Systems}
\author{Victoria E. Galanopoulou,~\IEEEmembership{Student Member,~IEEE}, Thrassos K. Oikonomou~\IEEEmembership{Student Member,~IEEE}, \\Odysseas G. Karagiannidis,~\IEEEmembership{Student Member,~IEEE},  Sotiris A. Tegos,~\IEEEmembership{Senior Member,~IEEE}, \\Panagiotis D. Diamantoulakis,~\IEEEmembership{Senior Member,~IEEE}
\thanks{
% V. E. Galanopoulou,  T. K. Oikonomou, O. G. Karagiannidis, S. A. Tegos,  and P. D. Diamantoulakis 
The authors are with the Aristotle University of Thessaloniki, 54124 Thessaloniki, Greece (e-mails: \{vgalanop, toikonom, okaragia, tegosoti, padiaman\}@auth.gr).}}

        % <-this % stops a space



\maketitle

\begin{abstract}
Pinching antennas enable dynamic control of electromagnetic wave propagation through reconfigurable radiating structures, but selecting an optimal subset of antennas remains a combinatorial problem with exponential complexity. This letter considers antenna subset selection for a waveguide-fed pinching antenna array serving ground users under a time-division access scheme. The achievable rate depends on the coherent superposition of the effective complex channel gains and is therefore highly sensitive to the relative phase alignment of the activated antennas. To address the prohibitive complexity of exhaustive search, we propose a Viterbi state selection (VSS) algorithm that exploits the phase structure of the combined received signal.
The trellis state is defined by a quantized representation of the phase of the accumulated complex gain, and a Viterbi-based survivor rule is used to prune dominated antenna subsets across stages.
Numerical results demonstrate that the proposed method achieves the same antenna selection and rate as exhaustive search, while reducing the computational complexity from exponential to polynomial in the number of available antennas.
\end{abstract}

\begin{IEEEkeywords}
Pinching antennas, antenna activation, trellis, complexity
\end{IEEEkeywords}
% \vspace{0.3in}
\section{Introduction}

Achieving multi-gigabit throughput in next-generation wireless systems requires highly directional transmission schemes that can overcome severe propagation losses while avoiding excessive radio-frequency hardware complexity. This challenge has motivated extensive research on millimeter-wave communication, large-scale antenna arrays, and reconfigurable intelligent surfaces \cite{ris1}. In this context, pinching antenna (PA) systems have recently emerged as a promising architecture that guides radio-frequency signals through a leaky dielectric waveguide and radiates energy only at selected pinch locations \cite{docomo2022pinching, sotiris_panos, pinch_principles,pinch_flexible,liu_tutorial}. By confining signal propagation inside the waveguide until radiation, PAs significantly reduce free-space attenuation while preserving a compact and cable-like physical structure. Activating specific pinches along the waveguide enables the base station to establish line-of-sight links with multiple indoor users using minimal analog circuitry, positioning PA arrays as a cost-effective and scalable solution for future high-frequency wireless deployments.

Most existing studies on PA systems have focused on coverage analysis, beam steering, and user scheduling under simplified transmission models \cite{ding_karag_downlink}. In particular, prior works commonly consider either a single active pinch or small clusters of closely spaced PAs positioned above the scheduled user, which enables simplified signal representations and tractable system analysis \cite{yuanwei_beamforming}. However, in practical low-complexity deployments, PAs are pre-placed at fixed locations along the waveguide to provide persistent line-of-sight connectivity without continuous tracking \cite{thrassos}. Under this constraint, performance enhancements rely on selecting an appropriate subset of the available pinches, making antenna selection a central design problem. The resulting combinatorial complexity renders exhaustive brute-force search infeasible for large arrays and has motivated recent works on greedy and heuristic antenna selection strategies for multi-pinch activation. More specifically, in \cite{fixed_pos_game_theory}, antenna activation with fixed pinching positions is addressed using a coalition-based game-theoretic framework, confirming that optimizing the activation pattern significantly improves the system rate, but converges to a Nash-stable solution driven by local utility improvements. Moreover, in \cite{ppga}, antenna activation in multi-feedpoint PA systems is addressed using a low-complexity projection-guided algorithm that exploits phase alignment, but the resulting solution strongly depends on the initial activation set and follows a single refinement trajectory, which limits its applicability to multi-user scenario. Finally, in \cite{ody_vik}, the same problem is addressed using deep learning-based optimization, which, once properly trained, drastically reduces computational complexity at the expense of optimality. To the best of the authors’ knowledge, no existing work provides a unified and scalable antenna selection methodology for PA systems that achieves brute-force-level performance with computationally efficient complexity. Such a methodology could be used to directly optimize discrete PA systems or to train machine learning models more efficiently, which is particularly important when the wireless environment and user requirements change over time.

In this work, we address the antenna activation problem in waveguide-fed PA arrays with fixed antenna positions, with the objective of maximizing the achievable rate under binary activation constraints. We propose the Viterbi state selection (VSS) algorithm  for antenna activation that takes advantage of the phase structure of the coherently combined received signal to efficiently explore the combinatorial activation space. The proposed approach significantly reduces the exponential complexity of exhaustive search to polynomial complexity while retaining near-optimal performance. The framework supports both single-user and multi-user transmission, since in the latter case  the channel characteristics of multiple users are jointly considered within a unified trellis-based selection process, in which each state consists of  a predefined vector of phases with cardinality equal to the number of users. Numerical results demonstrate that the proposed method achieves identical performance  to exhaustive brute-force search, even for large arrays, while operating at significantly lower computational complexity.

\section{System Model}

We consider a downlink communication scenario in which a base station (BS) is equipped with $N$ PAs, indexed by the set $\mathcal{N}=\{1,\ldots,N\}$, and serves $M$ single-antenna users indexed by $\mathcal{M}=\{1,\ldots,M\}$. A three-dimensional Cartesian coordinate system is adopted, where the users are randomly distributed over a square region on the $x$--$y$ plane with side length $L$. The position of the $m$-th user is denoted by $\boldsymbol{\psi}_m=(x_m,y_m,0)$, where $x_m$ and $y_m$ are independent random variables uniformly distributed over the interval $[-L/2,L/2]$. In the considered PA system and without loss of generality, the dielectric waveguide is assumed to be installed parallel to the $x$-axis, with its center aligned at $y=0$, a height of $H$ and a length of $L$, which corresponds to the square room side length. The PAs are uniformly deployed along the waveguide to reduce system complexity and ensure uniform spatial coverage.
The position of the $n$-th PA is given by
\begin{equation}
    \bm{\psi}^{\mathrm{Pin}}_n
    =
    \left(
    -\frac{L}{2} + \frac{(2n-1)L}{2N},\;
    0,\;
    H
    \right),
    \quad n = 1,\dots,N .
\end{equation}

\begin{figure}[h]
\centering
\begin{tikzpicture}[
  >=Latex,
  scale=0.8,
  transform shape,
  axis/.style={dashed, line width=0.6pt},
  dim/.style={<->, dotted, line width=0.6pt},
  plane/.style={line width=1pt},
  lab/.style={font=\footnotesize\normalfont},
  % slot (pinch) styles
  slot/.style={draw, line width=0.55pt, minimum width=0.9mm, minimum height=3.2mm, inner sep=0pt},
  slotOn/.style={slot, fill=gray},
  slotOff/.style={slot, fill=white}
]

% ====================
% Room (small, final size)
% ====================
\coordinate (A) at (-2.6,-1.2);
\coordinate (B) at ( 2.6,-1.2);
\coordinate (C) at ( 2.1, 1.0);
\coordinate (D) at (-3.1, 1.0);
\draw[plane] (A)--(B)--(C)--(D)--cycle;

% ====================

% ====================
% Height h
% ====================
\draw[dotted] (0,0) -- ++(0,2.4) node[midway,right] {$H$};

% ====================
% Cable as a very thin rectangle (same width as room)
% ====================
\def\cableY{2.30}     % bottom of cable
\def\cableT{0.16}     % cable thickness (thin)
\draw[line width=0.6pt, fill=white]
  (-2.6,\cableY) rectangle (2.6,\cableY+\cableT);

% ====================
% Pinches (slots) ABOVE the cable rectangle
% ====================
\def\slotY{\cableY+\cableT-0.06}  % center of slots above cable
\def\xStart{-2.45}
\def\xStep{0.22}

\foreach \i in {0,...,22} {
  \pgfmathsetmacro{\x}{\xStart + \i*\xStep}
  \pgfmathtruncatemacro{\p}{mod(\i,3)}
  \ifnum\p=0
    \node[slotOn] at (\x,\slotY) {};
  \else
    \node[slotOff]  at (\x,\slotY) {};
  \fi
}

% ====================
% Dimensions
% ====================
% +/- D/2 above cable
\draw[dim] (-2.6,\cableY+0.55) -- (2.6,\cableY+0.55);
\node[lab] at (-2.6,\cableY+0.78) {$-L/2$};
\node[lab] at ( 2.6,\cableY+0.78) {$+L/2$};

% +/- D below room
\draw[dim] (-2.6,-1.8) -- (2.6,-1.8);
\node[lab] at (-2.6,-2.05) {$-L/2$};
\node[lab] at ( 2.6,-2.05) {$+L/2$};

\node[anchor=south, inner sep=0pt] (BS)
  at (-3.5,\cableY-1)
  {\includegraphics[height=1.6cm]{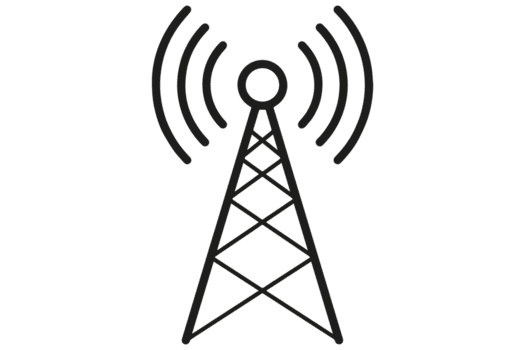}};
\node[anchor=center, inner sep=0pt] (User)
  at (0.4,-0.1)
  {\includegraphics[height=0.3cm]{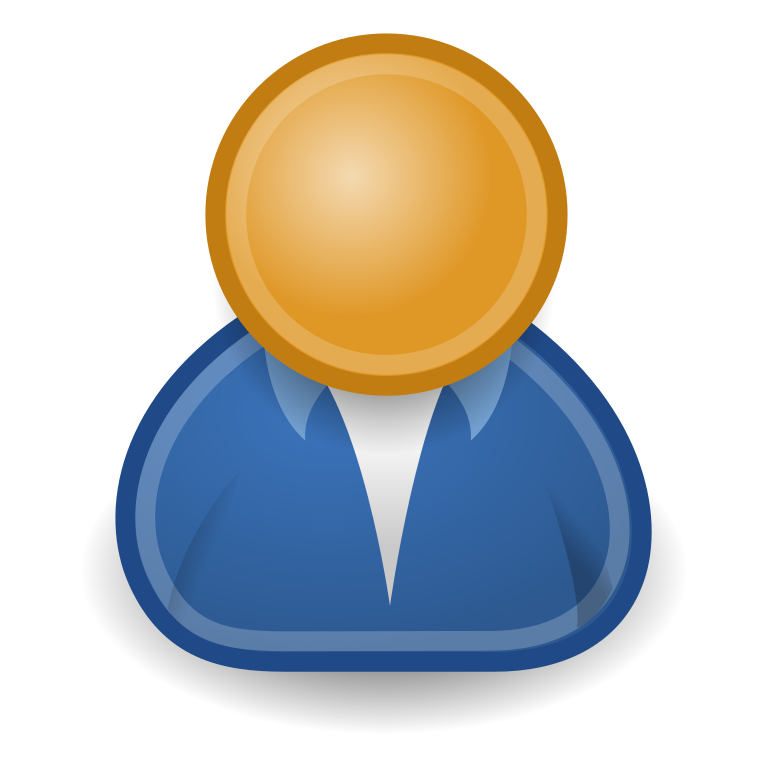}};
  \node[anchor=center, inner sep=0pt] (User)
  at (-0.8,0.1)
  {\includegraphics[height=0.3cm]{figures/user_icon.png}};

% ====================
% Legend
% ====================
\node[slotOn]  at (-3,-2.55) {};
\node[lab,right] at (-2.7,-2.55) {Activated PA};

\node[slotOff] at (1.1,-2.55) {};
\node[lab,right] at (1.35,-2.55) {Potential PA};

\end{tikzpicture}
\caption{System model.}
\end{figure}

The free-space channel between the $n$-th PA and the user is modeled as
\begin{equation}
    h_{m,n}
    =
    \frac{
    \exp\!\left(
    -j \frac{2\pi}{\lambda}
    \left\| \bm{\psi_m} - \bm{\psi}^{\mathrm{Pin}}_n \right\|
    \right)
    }{
    \left\| \bm{\psi_m} - \bm{\psi}^{\mathrm{Pin}}_n \right\|
    },
    \quad n = 1,\dots,N,
\end{equation}
where $\lambda$ denotes the carrier wavelength.
Since all $N$ PAs are deployed along the same waveguide, the signal radiated by each PA originates from a common feed and differs only by a deterministic phase shift introduced by propagation along the waveguide, i.e.,
\begin{equation}
    g_n = e^{-j{
    \frac{2\pi}{\lambda_g}
    \left\|
    \bm{\psi}^{\mathrm{Pin}}_n - \bm{\psi}^{\mathrm{Pin}}_0
    \right\|}},
\end{equation}
where $\bm{\psi}^{\mathrm{Pin}}_0$ denotes the position of the feedpoint of the waveguide, $\lambda_g = \lambda / n_{\mathrm{eff}}$ is the guided wavelength and $n_{\mathrm{eff}}$ denotes the effective refractive index of the dielectric waveguide.
For compact notation, the effective complex channel of the $n$-th PA is defined as 
\begin{equation}
B_{m,n} = h_{m,n} g_n, 
\end{equation}
and the corresponding vector of channels as $\mathbf{B_m} = [B_{m,1}, B_{m,2}, \dots, B_{m,N}]^{\mathsf T} \in \mathbb{C}^{N \times 1}$, where $(\cdot)^{\mathsf T}$ denotes the transpose operator.

We consider that only a subset of the available PAs is activated. Let $a_n \in \{0,1\}$ denote the binary activation variable associated with the $n$-th PA, and define the activation vector as
\(
\mathbf{a} = [a_1, a_2, \ldots, a_{N}]^{\mathsf T}
\).  The set of activated antennas is defined as
\begin{equation}
\mathcal{S} \triangleq \{\, n \in \{1,\ldots,N\}\mid a_n = 1 \,\}
\end{equation} 
with cardinality $|\mathcal{S}| = \|\mathbf{a}\|_0$. A time-division multiple access (TDMA) protocol is used so that only one user is served during each transmission interval. The received signal at each user is then expressed as
\begin{equation} \label{received}
     y_m
    =
    \sqrt{\frac{P\eta}{|\mathcal{S}|}}
    \mathbf{a}^{\mathsf T}\mathbf{B_m}\, s + w,
\end{equation}
where $\eta = \frac{c^2}{16\pi^2 f_c^2}$ is the free-space path-loss scaling factor, $c$ denotes the speed of light, $f_c$ the carrier frequency, $P$ the transmit power, $s \in \mathbb{C}$ the transmitted symbol, and $w \sim \mathcal{CN}(0,\sigma_w^2)$ the additive white Gaussian noise. It can be observed in \eqref{received} that the total transmit power is equally allocated among the activated antennas. Finally, the received signal-to-noise ratio (SNR) at the $m$-th user is given by
\begin{equation}
    \gamma_m
    =
    \frac{P\eta}{|\mathcal{S}| \sigma_w^2}
    \left|
    \mathbf{a}^{\mathsf T}\mathbf{B_m}
    \right|^2 
\end{equation}
and the achievable rate of the $m$-th user is expressed as
\begin{equation}
    R_m(\mathbf{a} ) = \log_2\!\left(1 + \gamma_m\right).
\end{equation}

\section{Optimization Problem}
In the considered PA system, the achievable rates of the different users are highly sensitive to the choice of the activated antenna set. However, due to the short duration of the transmission time slots, a common PA configuration is assumed to be used for all users. Each PA contributes a complex signal with a user-independent magnitude and phase so that a given PA may increase the received signal strength for certain users through constructive combining, but it may simultaneously degrade the performance of others. In addition, while activating more PAs may increase the received signal of a specific user through constructive combining, it also incurs a power-splitting penalty and phase misalignment effects. As a result, the achievable performance depends critically on the coherent superposition of these signals at the receivers.  
The objective is therefore to determine a common antenna activation set $\mathcal{S}$ that balances these conflicting effects across users by maximizing the minimum achievable rate, and equivalently the minimum received SNR, among all users.

This leads to the following worst-user (max-min) antenna activation problem
\begin{equation}
\label{eq:maxmin_snr}
\max_{\mathbf{a} \in \{0,1\}^N}
\;
\min_{m \in \{1,\ldots,M\}}
\frac{\left| \mathbf{a}^{\mathsf T}\mathbf{B}_m \right|^2}{\|\mathbf{a}\|_0}.
\end{equation}
In the special case of a single user, the max-min problem in \eqref{eq:maxmin_snr} reduces to
\begin{equation}
\max_{\mathbf{a} \in \{0,1\}^N}
\frac{\left| \mathbf{a}^{\mathsf T}\mathbf{B} \right|^2}{\|\mathbf{a}\|_0},
\end{equation}
which constitutes a quadratic fractional 0-1 programming (QF01P) problem and is nonlinear, nonconvex, and NP-hard \cite{ody_vik}.
As a result, exhaustive search over all antenna subsets becomes computationally prohibitive even for moderate $N$, due to the exponential growth of the search space, motivating the development of efficient structured algorithms such as the proposed VSS method.

\section{Viterbi State Selection}
\subsection{Proposed Algorithm}
The proposed VSS algorithm constructs a trellis that incrementally builds antenna activation sets by exploiting the phase structure of the effective channel gains and prunes suboptimal partial paths using a survivor selection rule. The trellis is organized according to the number of activated PAs and uses phase quantization to limit the number of candidate paths, following a Viterbi-like survivor selection principle. The conception of the proposed approach is based on the fact that for the same number of activated PAs, between two solutions that lead to the same phase, the one with the higher SNR can be selected. This is because a PA that is not selected at a specific phase can be selected at a later stage, while the phase of the accumulated phase, i.e., the phase of the superposed signal, from the different activated PAs can be seen as a form of memory. In more detail, a selection of a specific subset of active PAs only limits the accumulated phase and does not affect by any other means the decision to activate a PA that does not belong in the subset. More details on the definitions and the main steps of the proposed algorithm are provided below.

\paragraph{Stage definition}
The trellis consists of $N$ stages. Stage $\tau \in \{0,\dots,N\}$ corresponds to candidate activation sets S with exactly $\tau$ active antennas, i.e., activation vectors $\mathbf{a} \in \{0,1\}^N$ satisfying $\|\mathbf{a}\|_0=\tau$.

\paragraph{Effective signal accumulation}
For any candidate activation vector $\mathbf{a}$, the accumulated \emph{complex} signal for user $m$ is defined as
\begin{equation}
\label{eq:complex_sum}
Z_m(\mathbf{a})
\triangleq
\sum_{n=1}^{N} a_n B_{n,m}.
\end{equation}
The corresponding accumulated phase is then given by
\begin{equation}
\label{eq:phase_def}
\Phi_m(\mathbf{a}) \triangleq \angle\!\left( Z_m(\mathbf{a}) \right),
\end{equation}
where $\angle(\cdot)$ denotes the argument of a complex number. The joint accumulated phase vector is defined as
\begin{equation}
\boldsymbol{\Phi}(\mathbf{a})
\triangleq
\big[
\Phi_1(\mathbf{a}),\dots,\Phi_M(\mathbf{a})
\big].
\end{equation}

\begin{comment}
\paragraph{Effective signal accumulation}
%Recall that the effective complex gain of antenna $n$ is denoted by $B_n$.
For any candidate activation set
$\mathcal{S} \subseteq \{1,\dots,N\}$, we define the accumulated
\emph{complex} signal as
\begin{equation}
\label{eq:complex_sum}
Z_m(\mathcal{S})
\triangleq
\sum_{n \in \mathcal{S}} B_{m,n} .
\end{equation}
The corresponding accumulated phase is then given by
\begin{equation}
\label{eq:phase_def}
\Phi_m(\mathcal{S}) \triangleq \angle\!\left( Z_m(\mathcal{S}) \right),
\end{equation}
where $\angle(\cdot)$ denotes the argument of a complex number.
The corresponding phase state is then given by
\begin{equation}
\boldsymbol{\Phi}(\mathcal{S})
\triangleq
\big[
\Phi_1(\mathcal{S}),\dots,\Phi_M(\mathcal{S})
\big]
\end{equation}
\end{comment}

\paragraph{Quantized phase state}

Each accumulated phase $\Phi_m(\mathbf{a})$ is independently quantized into $Q$ uniform bins over the interval $[-\pi,\pi)$, with values from set \(\mathcal{Q}\triangleq
\{\mathcal{Q}_1,\ldots,\mathcal{Q}_Q\}\) given by
\begin{equation}
\label{eq:phase_bins}
\mathcal{Q}
\triangleq
\left\{
-\pi + \frac{(2k-1)\pi}{Q}
\right\}_{k=1}^{Q}.
\end{equation}
As a result, each activation vector $\mathbf{a}$ is mapped to a multi-dimensional trellis state
\begin{equation}
\boldsymbol{q}(\mathbf{a})
\triangleq
\big(q_1(\mathbf{a}), \ldots, q_M(\mathbf{a})\big),
\end{equation}
where $q_m(\mathbf{a})=\mathsf{Quant}(\Phi_m(\mathbf{a})) \in \mathcal{Q}$ denotes the quantized phase corresponding to user $m$, which assigns
$\Phi_m(\mathbf{a})$ to one of the $Q$ uniform phase bins. The total number of trellis states per stage is therefore $Q^M$.

\paragraph{Initialization (stage $\tau=0$)}
The trellis is initialized with a reference phase state corresponding to
zero accumulated phase. This initial state represents the absence of any active antenna, i.e., all-zero activation vector, and serves solely as a phase reference for subsequent
expansions, incurring no achievable rate contribution.
\begin{comment}
\paragraph{First expansion (stage $\tau=1$)}
From the reference state at $\tau=0$, the trellis is expanded by activating exactly one antenna at a time. Specifically, for every $n \in \{1,\dots,N\}$ we form the singleton set $\mathcal{S}=\{n\}$ and assign it to the corresponding multi-dimensional phase state $\mathbf{q'}$.  For each state $\mathbf{q'}$, where multiple PAs may be mapped, only the antenna yielding the highest minimum user rate is retained as the survivor at stage $\tau=1$, as detailed next.
\end{comment}
\paragraph{State transitions}
\begin{comment}
From the reference state at $\tau=0$, the trellis is expanded by activating exactly one PA at a time. Specifically, for every $n \in \{1,\dots,N\}$ we form vector $\mathbf{e}_n$ with one non-zero element that denotes that the $n$-th PA is activated.
% , and assign it to the corresponding multi-dimensional phase state $\boldsymbol{q}(\mathbf{a})$. 
For each destination state $\boldsymbol{q}$, where multiple PAs may be mapped, only the activation set yielding the highest minimum user rate is retained as the survivor at stage $\tau=1$, as detailed next.
\end{comment}
From the reference state at $\tau = 0$, the trellis is initialized by activating exactly one PA at a time. Specifically, for each $n \in \{1,\ldots,N\}$, an activation vector $\mathbf{e}_n$ is formed with one non-zero element indicating that the $n$-th PA is active. These initial activations are mapped to their corresponding discrete phase states, and for each destination state $\boldsymbol{q}$, only one activation pattern is kept as the survivor of stage $\tau = 1$, as detailed below.
For subsequent stages the trellis is evolved in a similar manner.
From a survivor activation pattern $\mathbf{a}_{\tau-1,\boldsymbol{q}}$ at stage $\tau-1$, new candidate activation patterns at stage $\tau$ are generated by activating one additional PA that is currently inactive. Specifically, for any index $n \in \{1,\ldots,N\}$ which corresponds to a deactivated antenna, a candidate activation vector is formed as
\begin{equation}
\label{eq:transition}
\mathbf{a}' = \mathbf{a}_{\tau-1,\boldsymbol{q}} + \mathbf{e}_n.
\end{equation}
% where $\mathbf{e}_n$ denotes the $n$-th canonical basis vector.
The accumulated complex signal for user $m$ is updated as
\begin{equation}
\label{eq:complex_update}
Z_m(\mathbf{a}')
=
Z_m(\mathbf{a}_{\tau-1,\boldsymbol{q}}) + B_{n,m},
\end{equation}
The corresponding worst-user SNR is then evaluated for the candidate pattern $\mathbf{a}'$. Only if this SNR exceeds that of the originating survivor $\mathbf{a}_{\tau-1,\boldsymbol{q}}$ is the candidate accepted and forwarded to the next stage. In that case, the new destination trellis state is obtained by phase quantization as \(\boldsymbol{q}(\mathbf{a}').\)

\paragraph{Survivors}
At each stage $\tau$ and for each multi-dimensional state $\mathbf{q}\in\mathcal{Q}^M$, the algorithm retains a single \emph{survivor} activation set $\mathcal{S}_{\tau,\mathbf{q}}$, equivalently represented by its activation vector $\mathbf{a}_{\tau,\mathbf{q}}$, defined as
\begin{equation}
\label{eq:survivor_def}
\mathbf{a}_{\tau,\mathbf{q}}
\in
\argmax_{\substack{\mathbf{a}\in\{0,1\}^N \\
\|\mathbf{a}\|_0=\tau,\; \mathbf{q}(\mathbf{a})=\mathbf{q}}}
M_{\tau,\mathbf{q}}(\mathbf{a}),
\end{equation}
where the survivor metric is given by
\begin{equation}
\label{eq:survivor_metric}
M_{\tau,\mathbf{q}}(\mathbf{a})
\triangleq
\min_{m \in \{1,\ldots,M\}}
\frac{\left| \mathbf{a}^{\mathsf T}\mathbf{B}_m \right|^2}{\|\mathbf{a}\|_0}.
\end{equation}
This one-survivor-per-state rule limits the number of candidate activation patterns at each stage to at most $Q^M$.
\begin{comment}
\paragraph{Survivors}
At each stage $\tau$ and for each multi-dimensional state $\mathbf{q'} $, the algorithm retains a single \emph{survivor} activation set $\mathcal{S}_{\tau,\mathbf{q'}}$, defined as
\begin{equation}
\label{eq:survivor_def}
\mathcal{S}_{\tau,\mathbf{q'}}
\in
\argmax_{\substack{\mathcal{S}\subseteq\{1,\dots,N\} \\
|\mathcal{S}|=\tau,\; \mathbf{q'}(\mathcal{S})=\mathbf{q'}}}
M_{\tau,\mathbf{q}}
\end{equation}
where $M_{\tau,\mathbf{q}}$ denotes the survivor metric, i.e.,
\begin{equation}
M_{\tau,\mathbf{q}} \triangleq \min_{m \in \{1,\ldots,M\}}
\frac{\left| \mathbf{a}^{\mathsf T}\mathbf{B}_m \right|^2}{\|\mathbf{a}\|_0}
\end{equation}

\end{comment}
\begin{comment}
\paragraph{State Transitions}
From a survivor $\mathcal{S}_{\tau-1,q'}$ at stage $\tau-1$, new candidates at stage $\tau$ are generated by activating one additional PA, besides the ones already in $\mathcal{S}_{\tau-1,q'}$, i.e.,
\begin{equation}
\label{eq:transition}
\mathcal{S}' = \mathcal{S}_{\tau-1,q} \cup \{a\},
\qquad
a \in \{1,\dots,N\} \setminus \mathcal{S}_{\tau-1,q}.
\end{equation}
The accumulated complex signal is updated as
\begin{equation}
\label{eq:complex_update}
Z(\mathcal{S}')
=
Z(\mathcal{S}_{\tau-1,q}) + B_a,
\end{equation}
and the destination state is obtained by phase quantization
\begin{equation}
\label{eq:state_update}
q' = \mathsf{Quant}\!\left( \Phi(\mathcal{S}') \right).
\end{equation}
\end{comment}
For each state $\boldsymbol{q}$, multiple candidate sets $\mathcal{S}'$ may be generated from different survivors. These candidates are compared and only the best one is retained as $\mathbf{a}_{\tau,\boldsymbol{q}}$ according to the Viterbi survivor rule.
\begin{comment}
\begin{equation}
\label{eq:viterbi_update}
\text{if } R(\mathcal{S}') > M_{\tau,q'} \text{ then set }
(\mathcal{S}_{\tau,q'}, M_{\tau,q'}) \leftarrow (\mathcal{S}', R(\mathcal{S}')).
\end{equation}
\end{comment}
\paragraph{Iteration and termination}
The above procedure is repeated for increasing stages $\tau$ until either all $N$ stages are processed or no candidate extension yields an improvement in any survivor path, in which case the algorithm terminates early.

\paragraph{Final selection}
The final antenna activation pattern is selected as the best survivor across all survivor paths:
\begin{equation}
\label{eq:final_pick}
\mathbf{a}^\star
\in
\argmax_{\tau\in \{1,\dots,T\},\;
\boldsymbol{q} \in \mathcal{Q}^M}
M_{\tau,\boldsymbol{q}},
\end{equation}
where $T$ denotes the last executed stage.

\begin{figure}[h]{
\centering
\includegraphics[width=0.95\columnwidth]{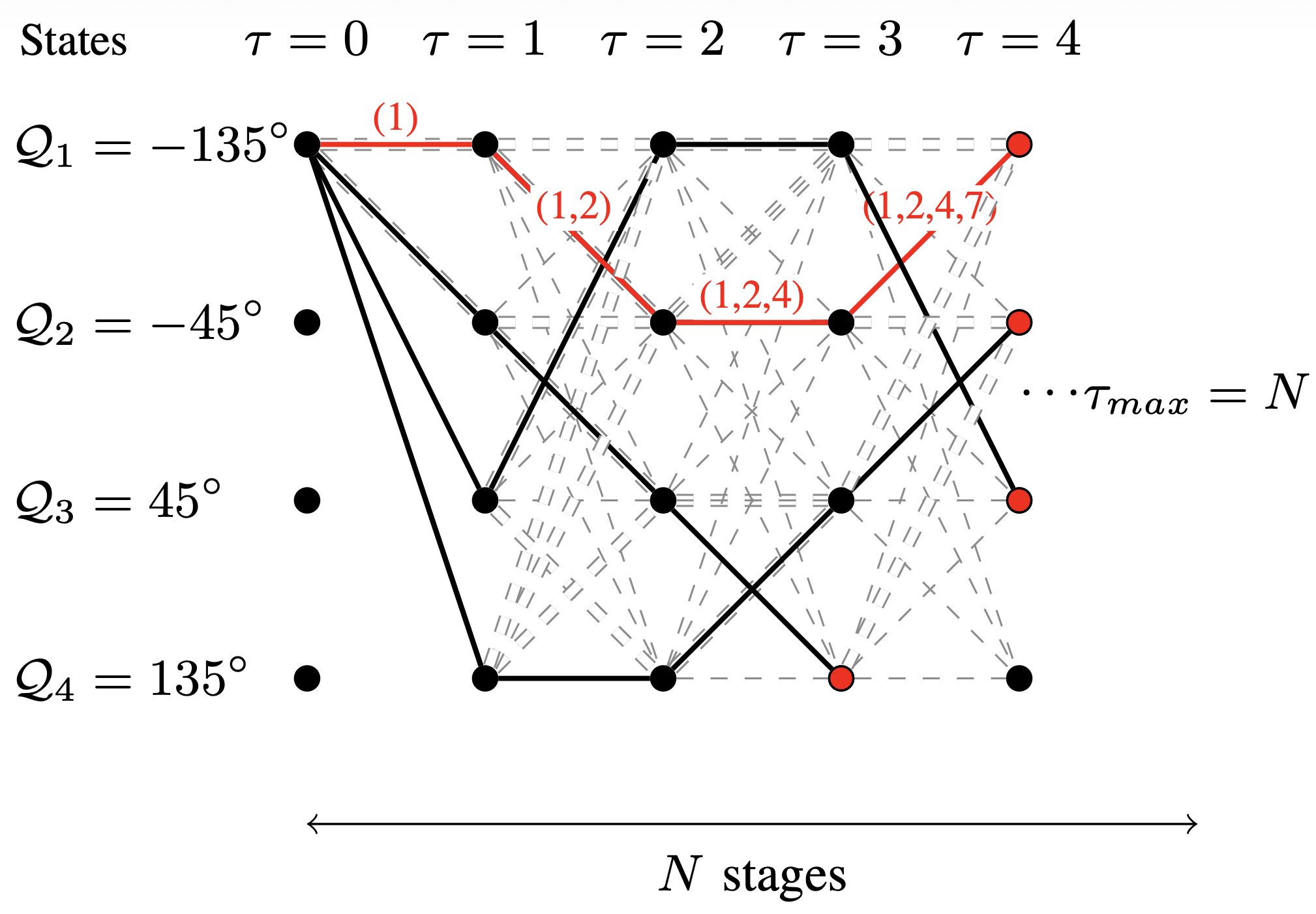}
}
\caption{Graph representation of the trellis  algorithm ($Q=4$, $N=8$, $M=1$).}
\label{fig:pinching_graph}
\end{figure}

In Fig. \ref{fig:pinching_graph}, the trellis diagram of the algorithm is presented. We consider the case of a single user, for which the states correspond to the discrete phase states \(\mathcal{Q}\) with \(Q = 4\) states, and total number of PAs \(N = 8\). At each time stage $\tau$, multiple candidate transitions are evaluated and a survivor is selected for each state (solid black edges) as described above, forming a trellis maximum $N$ stages. We consider that the algorithm stopped earlier at \(\tau=4\), as no further improvement is observed with any additional antenna. Each path through the trellis represents a candidate solution, while the solid red path denotes the survivor trajectory selected by the algorithm. The labels  (numbers in parentheses) denote the active antennas associated with the survivor transitions along the chosen path.

\subsection{Computational Complexity}

The computational complexity of the proposed algorithm is governed by the number of trellis states and candidate extensions evaluated at each stage. In the single-user case, each stage contains at most $Q$ survivor states and each survivor can be extended by at most $N$ remaining antennas. Since there are at most $N$ stages, the overall complexity scales as $\mathcal{O}(Q N^2)$ metric evaluations. This bound is conservative, as it assumes that all $N$ stages are fully explored and that all $N$ antennas are examined at each stage, whereas in practice only $N-t$ candidate extensions are considered at stage $t$, and the trellis typically terminates well before all stages are visited, as confirmed by the numerical results in Section~\ref{s_nr}.
In the multi-user case, the number of trellis states increases to $Q^M$ and the complexity correspondingly becomes $\mathcal{O}(Q^M N^2)$. In contrast, the brute force solution requires an exhaustive search over all non empty antenna subsets, resulting in an exponential complexity of $\mathcal{O}(2^N)$ metric evaluations. Therefore, the proposed trellis-based approach replaces the exponential dependence on $N$ with a polynomial complexity while retaining a structured exploration of the solution space, making it computationally tractable even for moderately large antenna arrays.

\section{Numerical Results} \label{s_nr}

In this section, numerical simulations are conducted to evaluate the performance of the proposed algorithm.
The waveguide is placed at a height of $H = 3$~m and has a length of
$L = 50$~m. The carrier frequency is set to $f_c = 28$~GHz, the noise
power is $\sigma^2 = -90$~dBm, and the effective refractive index of the waveguide is $n_{\mathrm{eff}} = 1.4$. All reported results are averaged over 1000 random user locations, uniformly distributed over the considered room area. For the proposed VSS algorithm, we set \(Q=4\). Although increasing 
$Q$ allows the algorithm to retain multiple phase-consistent activation paths, we observe that the achievable performance saturates quickly with 
$Q$, and small values (e.g., $Q=4$) are sufficient to attain near-optimal performance across the considered scenarios.

\begin{comment}
In the extreme case $Q=1$, the framework reduces to a greedy forward-selection procedure, which achieves competitive performance in many realizations. However, it lacks robustness in unfavorable channel realizations, particularly in the multi-user max-min setting. 
\end{comment}

For \(N \leq 20\), the optimal benchmark is obtained via exhaustive search over all antenna subsets. 
For larger \(N\), a commercial mixed-integer optimization solver, Gurobi, is used and is guaranteed to converge to the global optimum, although it is computationally expensive~\cite{ody_vik}.
\begin{comment}
For larger \(N\), Gurobi optimizer is used: a Dinkelbach-based approach is applied in the single-user case, while the multi-user max-min problem is solved via an epigraph formulation (Aprendix A). 
These methods are computationally expensive but guaranteed to converge to the global optimum~\cite{ody_vik}. 
\end{comment}
As an additional benchmark, we consider the projection-guided greedy activation (PGGA) algorithm \cite{ppga}, originally proposed for multi-feedpoint and single user PA systems, and adapt it to the single-feedpoint setting considered here.

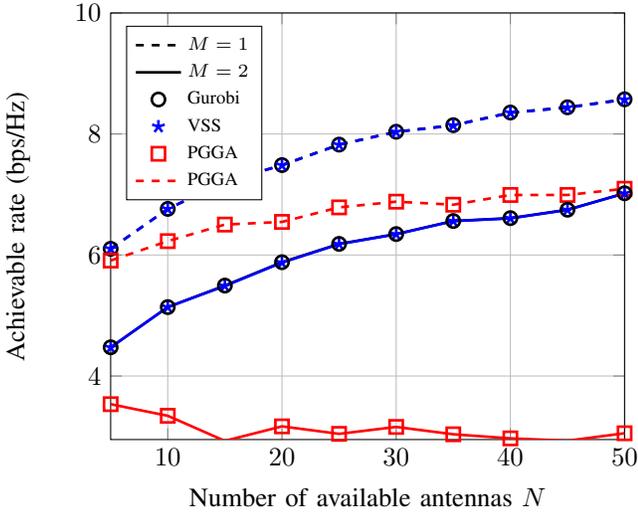
\begin{figure}[h]
\centering
\begin{tikzpicture}
    \begin{axis}[
        width=0.95\linewidth,
        xlabel={Number of available antennas $N$},
        ylabel={Achievable rate (bps/Hz)},
        ymin=2.95,
        ymax=10,
        xmin=5.0,
        xmax=50,
        grid=major,
        legend cell align = {left},
        legend pos = north west,
        legend style={font=\scriptsize}
    ]
    % Add plots
    % \addplot[black, mark=o, line width=1pt]
    % table {figures/rate_gurobi.dat};
    % \addlegendentry{Gurobi}
    \addplot[
            color= black,
            no marks,
            line width = 1pt,
		      style = dashed,
            mark=o,
            mark repeat = 1,
            mark size = 2.5,
            ]
            table {figures/rate_gurobi.dat};
            \addlegendentry{$M=1$}

    \addplot[
            color= black,
            no marks,
            line width = 1pt,
		      style = solid,
            mark=o,
            mark repeat = 1,
            mark size = 2.5,
            ]
            table {figures/gurobi_multiuser_final_worst_user_rate_vs_N_U2.dat};
            \addlegendentry{$M=2$}

    \addplot[
            color= black,
            only marks,
            line width = 1pt,
		      style = solid,
            mark=o,
            mark repeat = 1,
            mark size = 2.5,
            ]
            table {figures/rate_gurobi.dat};
            \addlegendentry{Gurobi}

    \addplot[
            color= blue,
            only marks,
            line width = 1pt,
		      style = solid,
            mark=star,
            mark repeat = 1,
            mark size = 2.5,
            ]
            table {figures/rate_trellis.dat};
            \addlegendentry{VSS}

    \addplot[
            color = red,
            only marks,
            line width = 1pt,
		      style = solid,
            mark=square,
            mark repeat = 1,
            mark size = 2.5,
            ]
            table {figures/rate_pgga.dat};
            \addlegendentry{PGGA}
    \addplot[
            color = red,
            no marks,
            line width = 1pt,
		      style = dashed,
            mark=square,
            mark repeat = 1,
            mark size = 2.5,
            ]
            table {figures/rate_pgga.dat};

    \addplot[
            color= blue,
            no marks,
            line width = 1pt,
		      style = dashed,
            mark=star,
            mark repeat = 1,
            mark size = 2.5,
            ]
            table {figures/rate_trellis.dat};
    %%%%% M=2
    \addplot[
            color= blue,
            only marks,
            line width = 1pt,
		      style = solid,
            mark=star,
            mark repeat = 1,
            mark size = 2.5,
            ]
            table {figures/trellis_multiuser_final_worst_user_rate_vs_N_U2.dat};

    \addplot[
            color= blue,
            no marks,
            line width = 1pt,
		      style = solid,
            mark=star,
            mark repeat = 1,
            mark size = 2.5,
            ]
            table {figures/trellis_multiuser_final_worst_user_rate_vs_N_U2.dat};
    \addplot[
            color= black,
            only marks,
            line width = 1pt,
		      style = solid,
            mark=o,
            mark repeat = 1,
            mark size = 2.5,
            ]
            table {figures/gurobi_multiuser_final_worst_user_rate_vs_N_U2.dat};       
    \addplot[
            color = red,
            only marks,
            line width = 1pt,
		      style = solid,
            mark=square,
            mark repeat = 1,
            mark size = 2.5,
            ]
            table {figures/pgga_multiuser.txt};
            \addlegendentry{PGGA}
     \addplot[
            color = red,
            no marks,
            line width = 1pt,
		      style = solid,
            mark=square,
            mark repeat = 1,
            mark size = 2.5,
            ]
            table  {figures/pgga_multiuser.txt};

    \end{axis}
\end{tikzpicture}
\caption{Comparison of achievable rates.}
\label{fig3}

\end{figure}

Fig.~\ref{fig3} illustrates the achievable rates obtained by the optimal solution, the VSS method, and the PGGA benchmark as a function of the number of available PAs.
The simulation results demonstrate that the VSS algorithm matches the optimal performance in both single and multi-user scenarios,
while reducing significantly the search complexity from exponential in $N$ to a polynomial-order procedure. 
It is also shown that the VSS approach consistently outperforms the PGGA benchmark in the considered scenarios. This is justified given that PGGA was originally designed for multi-feedpoint systems, where each PA can be assigned to different feedpoints to realize discrete phase control, and, in the single-feedpoint setting, relies solely on natural phase alignment without joint phase refinement. As a result, its performance becomes increasingly limited in the multi-user scenario, where a common antenna configuration should simultaneously accommodate multiple, potentially conflicting, phase-alignment requirements.

\begin{comment}
In Fig.~\ref{fig4}, we present the stage at which the proposed trellis algorithm terminates to illustrate that traversing all $N$ stages is generally unnecessary. Early termination occurs when no further antenna activation yields an improvement over any existing survivor path, which effectively determines the number of antennas activated by the algorithm.
To further characterize this behavior, we compare the termination stage $T_{\mathrm{stop}}$ with the stage $T_{\mathrm{best}}$ at which the maximum achievable rate is obtained. Since the final solution is selected as the best candidate across all previously explored stages, the optimal activation set may correspond to a survivor path that was identified at an earlier stage and remains unmatched in subsequent expansions.
As observed in the results, $T_{\mathrm{stop}}$ is significantly smaller than the total number of available stages, demonstrating that the trellis exploration typically terminates well before reaching the final layer, reducing further the complexity from the upper bound of $\mathcal{O}(Q N^2)$. 
Moreover, $T_{\mathrm{best}}$, which directly reflects the number of active antennas, is slightly smaller than
$T_{\mathrm{stop}}$, indicating that the optimal solution subset is found near the point of termination and confirming that is often identified before the algorithm terminates. 
\end{comment}

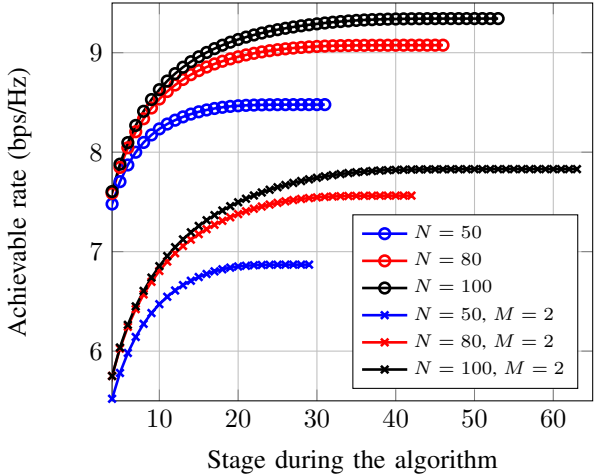
\begin{figure}[h]
\centering
\begin{tikzpicture}
    \begin{axis}[
        width=0.9\linewidth,
        xlabel={Stage during the algorithm},
        ylabel={Achievable rate (bps/Hz)},
        ymin=5.5,
        ymax=9.5,
        xmin=4,
        xmax=65,
        grid=major,
        legend entries = {
            {$N=50$},
            {$N=80$},
            {$N=100$},
            {$N=50$, $M=2$},
            {$N=80$, $M=2$},
            {$N=100$, $M=2$}
        },
         legend cell align = {left},
        legend pos = south east,
        legend style={
        % at={(1,1)},
        % anchor=north east,
        font = \scriptsize
        }   
    ]
    \addplot[blue,  mark=o, line width=1pt]
    table {figures/trellis_avg_rate_vs_stage_N50_users150.dat};
    \addplot[red,  mark=o, line width=1pt]
    table {figures/trellis_avg_rate_vs_stage_N80_users150.dat};
    \addplot[black,  mark=o, line width=1pt]
    table {figures/trellis_avg_rate_vs_stage_N100_users150.dat};
    \addplot[blue,  mark=x,line width=1pt]
    table {figures/trellis_multiuser_avg_maxmin_vs_stage_N50_U2.dat};
    \addplot[red,  mark=x,line width=1pt]
    table {figures/trellis_multiuser_avg_maxmin_vs_stage_N80_U2.dat};
    \addplot[black,  mark=x,line width=1pt]
    table {figures/trellis_multiuser_avg_maxmin_vs_stage_N100_U2.dat};
    \end{axis}
\end{tikzpicture}
\caption{Convergence of the trellis-based algorithm during stages.}
\label{fig5}
\end{figure}

In addition, Fig.~\ref{fig5} illustrates the convergence of the achievable rate across the trellis stages for different number of available PAs ($N = 50$, $N = 80$, $N=100$) and for both single and multi-user case. We present the stage at which the proposed VSS algorithm terminates to illustrate that traversing all $N$ stages is generally unnecessary. As observed in the results, the stage when the algorithm terminates is significantly smaller than the total number of available stages(number of PAs), demonstrating that the trellis exploration typically terminates well before reaching the final layer, reducing further the complexity from the upper bound of $\mathcal{O}(Q^M N^2)$. Moreover, it is shown that the rate typically saturates well before the termination stage is reached, indicating that further stage expansions yield negligible performance improvement. This observation suggests that the VSS algorithm could, in principle, be stopped even earlier based on a convergence criterion on the achievable rate, leading to an additional reduction in computational complexity.

\begin{comment}
\begin{figure}[h]
\centering
\begin{tikzpicture}
    \begin{axis}[
        width=0.9\linewidth,
        xlabel={Number of available antennas},
        ylabel={Stages},
        ymin=2,
        ymax=20,
        xmin=4,
        xmax=55,
        grid=major,
        legend entries = {
            {Tstop},
            {Tbest}
        },
        legend cell align = {left},
        legend pos = south east,
    ]
    \addplot[red,  mark=x,line width=1pt]
    table {figures/iters_Tstop.dat};
    \addplot[black, mark=x, line width=1pt]
    table {figures/iters_Tbest.dat};
    \end{axis}
\end{tikzpicture}
\caption{Termination stage $T_{\mathrm{stop}}$ and best stage $T_{\mathrm{best}}$ of the trellis-based algorithm as a function of the number of available PAs.}
\label{fig4}
\end{figure}
\end{comment}

\section{Conclusion}
In this work, we addressed the antenna activation problem in waveguide-fed PA systems with fixed antenna locations for single and multi-user scenario. To overcome the exponential complexity of exhaustive search, we proposed a VSS algorithm that exploits the phase structure of the accumulated received signal through a quantized phase-state representation to find the optimal antenna configurations. By retaining only one survivor per phase state and stage, the proposed method reduces the search complexity from exponential to polynomial order while preserving a structured exploration of the activation space. 
Numerical results demonstrate that the proposed VSS algorithm attains near-optimal achievable rates at substantially lower computational complexity than Gurobi solver, while significantly outperforming the PGGA benchmark.
Furthermore, the observed early termination and rapid rate convergence across trellis stages indicate that the practical computational complexity is substantially lower than the \(\mathcal{O}(Q^M N^2)\) worst-case complexity bound, explaining why exhaustive exploration of all antenna combinations is unnecessary in realistic scenarios.

\begin{comment}
\appendices
\section{}
This appendix provides an equivalent epigraph-based formulation of the worst-user antenna activation problem in \eqref{eq:maxmin_snr} and clarifies its relation to QF01P.
Starting from
\begin{equation}
\label{eq:maxmin_snr_app}
\max_{\mathbf{a} \in \{0,1\}^N}
\;
\min_{m \in \{1,\ldots,M\}}
\frac{\left| \mathbf{a}^{\mathsf T}\mathbf{B}_m \right|^2}{\|\mathbf{a}\|_0},
\end{equation}
an equivalent epigraph form can be obtained by introducing an auxiliary variable $t$ that lower bounds the minimum user performance, yielding
\begin{equation}
\label{eq:epigraph_app}
\begin{aligned}
\max_{\mathbf{a} \in \{0,1\}^N,\; t} \quad & t \\
\text{s.t.} \quad &
\frac{\left| \mathbf{a}^{\mathsf T}\mathbf{B}_m \right|^2}{\|\mathbf{a}\|_0}
\ge t, \quad m = 1,\ldots,M .
\end{aligned}
\end{equation}
Since $\mathbf{a}$ is binary, the numerator can be expressed as a quadratic form
\begin{equation}
\left| \mathbf{a}^{\mathsf T}\mathbf{B}_m \right|^2
=
\mathbf{a}^{\mathsf T}\mathbf{Q}_m \mathbf{a},
\qquad
\mathbf{Q}_m \triangleq \Re\!\left( \mathbf{B}_m \mathbf{B}_m^{\mathsf H} \right),
\end{equation}
while the denominator satisfies $\|\mathbf{a}\|_0 = \mathbf{1}^{\mathsf T}\mathbf{a}$.
Accordingly, each constraint in \eqref{eq:epigraph_app} can be rewritten as
\begin{equation}
\mathbf{a}^{\mathsf T}\mathbf{Q}_m \mathbf{a}
\;\ge\;
t\, \mathbf{1}^{\mathsf T}\mathbf{a},
\qquad m = 1,\ldots,M,
\end{equation}
which is a quadratic fractional constraint with binary variables.

\end{comment}

\bibliographystyle{IEEEtran}
\bibliography{bib}

\end{document}